\documentclass[conference]{IEEEtran}

\IEEEoverridecommandlockouts

\usepackage{cite}
\usepackage{amsmath,amssymb,amsfonts}
\usepackage{algorithmic}
\usepackage{graphicx}
\usepackage{textcomp}
\usepackage{xcolor}
\usepackage{booktabs} 
\usepackage{multirow}
\usepackage{subcaption} 
\usepackage{comment}

\def\BibTeX{{\rm B\kern-.05em{\sc i\kern-.025em b}\kern-.08em
    T\kern-.1667em\lower.7ex\hbox{E}\kern-.125emX}}
\begin{document}

\title{Learning Marmoset Vocal Patterns with a Masked Autoencoder for Robust Call Segmentation, Classification, and Caller Identification\\
\thanks{This work was supported by RIKEN Center for Advanced Intelligence Project (AIP), Japan.}
}

\author{\IEEEauthorblockN{Bin Wu\textsuperscript{1}, Shinnosuke Takamichi\textsuperscript{1,2}, Sakriani Sakti\textsuperscript{1,3}, Satoshi Nakamura\textsuperscript{1,3,4}}
\IEEEauthorblockA{\textsuperscript{1}RIKEN Center for Advanced Intelligence Project (AIP), Japan\\
\textsuperscript{2}Keio University, Japan\\
\textsuperscript{3}Nara Institute of Science and Technology, Japan\\
\textsuperscript{4}The Chinese University of Hong Kong, Shenzhen, China}
}



\maketitle

\begin{abstract}
  The marmoset, a highly vocal primate, is a key model for studying social-communicative behavior. Unlike human speech, marmoset vocalizations are less structured, highly variable, and recorded in noisy, low-resource conditions. Learning marmoset communication requires joint call segmentation, classification, and caller identification—challenging domain tasks. Previous CNNs handle local patterns but struggle with long-range temporal structure. We applied Transformers using self-attention for global dependencies. However, Transformers show overfitting and instability on small, noisy annotated datasets. To address this, we pretrain Transformers with MAE—a self-supervised method reconstructing masked segments from hundreds of hours of unannotated marmoset recordings. The pretraining improved stability and generalization. Results show MAE-pretrained Transformers outperform CNNs, demonstrating modern self-supervised architectures effectively model low-resource non-human vocal communication.
\end{abstract}


\section{Introduction}
The common marmoset (Callithrix Jacchus) is an animal model suitable for studying social vocal communication. First, the marmosets are non-human primates genetically and neurophysiologically close to human~\cite{kaas2016evolution}. Second, in contrast to such primates as gorillas that primarily use gestures for communication, the marmosets are highly vocal and readily to respond to other marmosets, even non-related or non-pair-bonded ones, particularly when visually hindered such as in forests when vocal contact is crucial for survival~\cite{chen2009contact}. Third, marmosets exhibit human-like conversational turn-taking, exchanging calls resembling coupled oscillators~\cite{takahashi2013coupled,wilson2005oscillator}. Fourth, marmosets display prosocial behaviors: similar to human cooperative breeding, marmosets take care of offspring of nonparents~\cite{burkart2009cooperative,burkart2010cognitive}, reflecting group social communications.

Researchers have been using the marmoset as an animal model to study diseases and mechanisms related to vocal communication comparing with human infant linguistic developments. Uesaka et al.~\cite{uesaka2023classification} developed a marmoset model of autism disorder --- marked by the deficits in social communication, impaired verbal interaction, and verbal perseveration --- by feeding pregnant mothers with valproic acid. Using the autism model, they aimed to study the developmental vocal characteristics of autism for early diagnosis and investigate potential medicines to relieve the autism symptoms. The ability of vocal communication is shaped by innate and empirical factors. Researchers manipulate autism-related genes of marmosets~\cite{kishi2014common} to study innate gene-function relationships. Researchers manipulate environments including visual or auditory sensory inputs~\cite{liao2018internal} and parental interaction~\cite{takahashi2015developmental} to study the empirical modification of communicative behaviors.

Studying turn-taking vocal communication between marmosets requires extracting caller and callee information, call contents, and vocal exchanges from the recorded audios. Turesson et al. applied SVM and DNN for marmoset call classification on a small dataset of 321 marmoset calls~\cite{turesson2016machine}. Wisler et al. used SVM and decision tree on a larger dataset of 4 call types, each with 400 marmoset calls for classfication~\cite{wisler2016framework}. Uesaka et al. employed CNN to classify 3 call types (phee, twitter, and trill) to study the development and autism of the marmosets~\cite{uesaka2023classification}. However, these studies were limited by either small datasets or focused on only a few call types.

Zhang et al. applied RNN and DNN for segmentation and classification of infant marmoset calls on a dataset that contains 10 call types, each with several thousand calls~\cite{zhang2018automatic}. Their call types include phee, twitter, trill, trillphee, tsik, ek, pheecry, peep, and two infant-specific categories: ct-trill (twitter-connected-trill) and ct-phee (twitter-connected-phee). Sarkar et al. used the same dataset and implemented self-supervised learning on caller discrimination and classification~\cite{sarkar2023can}. These studies have two key limitations. First, both studies ~\cite{zhang2018automatic,sarkar2023can} treat segmentation as a separate task and assumed known segmentation information when classifying calls or callers.
Second, their dataset recorded individual marmosets in isolation, without communicative interaction with other marmosets, thus failing to capture the vocal characteristics of marmosets during social communication.

Oikarinen et al. used a dataset of recordings from paired of pairs marmosets housed together in single cages, enabling close-range vocal interaction~\cite{landman2020close,oikarinen2019deep}. The dataset comprises 36 sessions totally 38 hours, with 8 call types: trill, phee, trillphee, twitter, chirp, tsik, ek, and chatter, along with a noise type that indicates the silences between the calls. Applying a CNN model on the dataset, they achieved segmentation, classification, and caller identification~\cite{oikarinen2019deep}.

However, while previous works as~\cite{zhang2018automatic} and~\cite{oikarinen2019deep} used an RNN or a CNN that maps acoustic segments to call labels, the Transformer structure has been proven to outperform RNN~\cite{vaswani2017attention} and CNN~\cite{dosovitskiy2021image} in both sequential language processing and high-dimensional vision tasks. Transformer utilizes self-attention mechanism that efficiently segregates information parallelly over long distances and captures the global structure of marmoset vocalization more efficiently than RNN and more effectively than CNN. While Transformer is typically constrained by the quadratic complexity of input token length~\cite{vaswani2017attention}, making it challenging to process high-dimensional input spectra that discriminate the marmoset calls such as a phee and a trill. We can address the limitation using Vision Transformer~\cite{dosovitskiy2021image} that patchizes the high-dimensional input. We propose to use the Transformer model for segmentation, classification, and caller identification of marmoset vocalizations.

However, Transformer shows the problems of overfitting and unstable training on the target dataset. The available annotated two-stream dataset~\cite{landman2020close} specifically designed for caller annotation has a limited amount. We found that typical 12-layer Transformer overfits on the dataset just as CNN does. We have to decrease both the number of layers and model dimension to reduce overfitting. The complex task that involves segmentation, classification, and caller identification on noisy data also challenges Transformer training. We observed sudden drops in accuracy, some of which never recover to the original level. Such training instability is more severe when we use larger Transformer models.

To address the problem of the overfitting and unstable training of large Transformer with limited annotated data on challenging tasks, we propose using Masked Autoencoder (MAE)~\cite{He2022} to pretrain the Transformer. We applied MAE on hundreds of hours of marmoset recordings without any annotations. We found that after pretraining, the Transformer shows more stable training and almost no overfitting on the limited dataset. We frozed lower layers to extract MAE features and used higher layers for fine-tuning, enabling us to use a typical 12-layer Transformer without reducing model dimension. The MAE pretrained Transformer works best in practice.

We conducted our experiments on the public dataset~\cite{landman2020close} with recordings designed to study close-range vocal communication between marmoset pairs who exchange calls close together in a cage.

\begin{figure}[t]
  \centering
  \includegraphics[width=\linewidth]{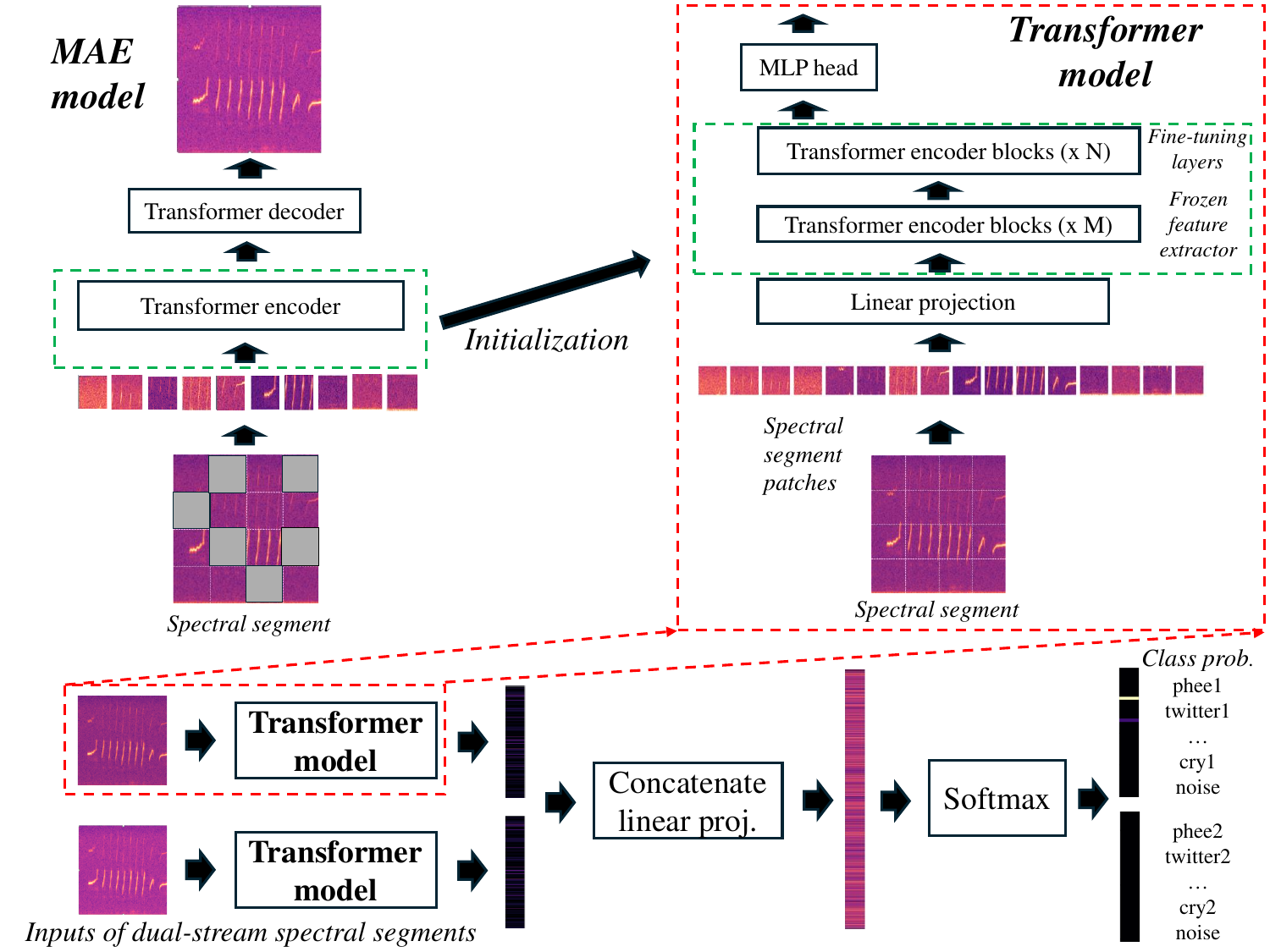}
  \caption{The two-stream Transformer model. The Transformer model can be initialized with MAE encoder pretrained by spectral segments without labels. We can freeze the first N layers for MAE feature extraction and the later N layers for fine-tuning. Two Transformer models process two spectral segments extracted from a dual-channel audio recording to map to a vocal call label. The call label of 'phee2' denotes that the phee call comes from the second marmoset of the pair in the cage; the 'noise' denotes the non-call segment between calls, used for segmenting calls to non-calls. We used two output heads to enable simultaneous call prediction, allowing detection of concurrent calls of the animal pair such as 'trill1' and 'twitter2'.}
  \label{fig:model}
\end{figure}

\section{Model}
We use a two-stream Transformer model (Figure~\ref{fig:model}) on the dual-audio recordings with two simultaneously recorded channels that come from two interacting animals. The model employs two stream transformer encoders that process sliding spectral segments from each channel to classify into a target label. The target labels indicate call types and caller identities (e.g., the 'tr2' label denotes a trill call from the second of two animals in interactions). The labels also indicate segmentation information: 'noise' denotes the non-call segment between calls.
This approach allows our two-stream Transformer model to perform three crucial tasks:
\begin{itemize}
    \item \textbf{Segmentation}: Identifying the start and end times of each vocalization.
    \item \textbf{Classification}: Determining the type of call (e.g., trill, chirp).
    \item \textbf{Caller identification}: Attributing each call to the correct animal.
\end{itemize}
By integrating these functions into a single model, we capture the complex dynamics of animal interactions through their vocalizations.

For our two-stream Transformer model, we utilize two Vision Transformer~\cite{vaswani2017attention} modules; each (Figure~\ref{fig:model}) processes the high-resolution linear spectral image by dividing it into patches that form a sequence of patch tokens. These tokens undergo a linear transformation and are augmented with positional encoding and a learnable class token. The resulting sequence passes through a typical Transformer architecture that comprises alternated self-attention and feedforward modules. The two-stream Transformer's vision transformer modules output two encoded class tokens, which are then linear projected and concatenated, passing through a shared linear layer for final class label prediction.

We can initialize each stream of Vision Transformer with identical parameters from the SSL model of MAE pretrained on spectral segments without labels. We froze the first M layers for MAE feature extraction and the later N layers for task-specific fine-tuning. This method retains the generalizability of MAE features while fine-tuning fewer parameters, making training faster and reducing overfitting risk for small annotated datasets.

\begin{figure}[t]
  \centering
  \includegraphics[width=\linewidth]{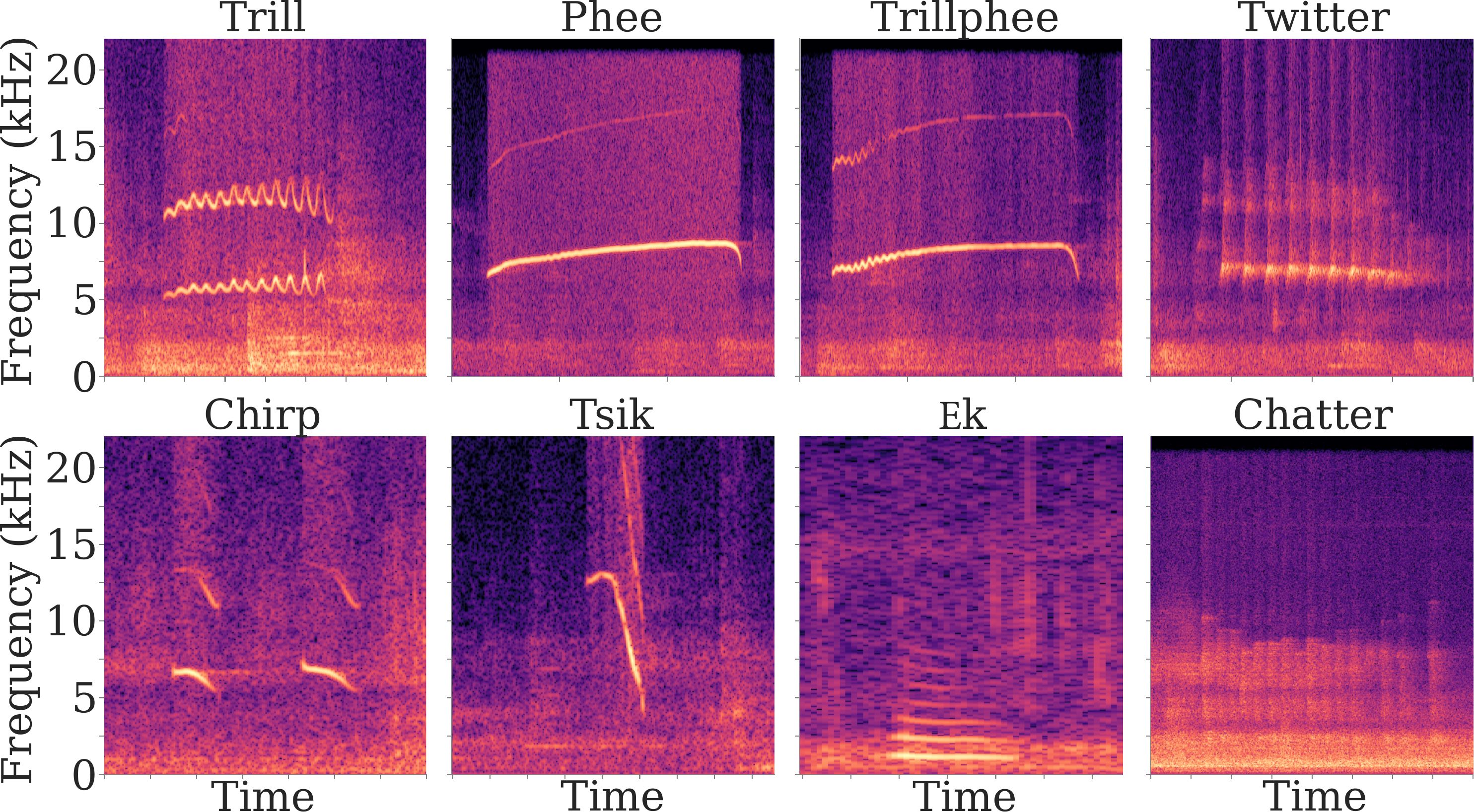}
  \caption{Spectrograms of marmoset calls. Note that ranges in time axis vary greatly among calls, e.g., 0.01~second for ek and 4~seconds for chatter. Those call examples come from the dataset~\cite{landman2020close}. We experimented on this dataset with the 8 call types, same as~\cite{oikarinen2019deep}.}
  \label{fig:call examples_takamichi}
\end{figure}  

\section{Evaluation}
We use F-score and accuracy to evaluate classification, segmentation, and caller identification, same as~\cite{oikarinen2019deep}.

For caller evaluation, we convert predicted labels into two separate segment files for a marmoset pair (e.g., 'tr2' labels are converted to 'tr' and added to the segment file of the second marmoset).We evaluate two predicted segment files against corresponding annotated ones for marmoset pair housed together.

For segmentation evaluation, we process the annotated and predicted segment files. First, we add 'noise' labels to fill in the intervals without any calls in the annotated segment files. Second, we reconstruct the interval for each call by merging predicted successive identical labels in the predicted segment files (e.g., the predicted 'noise, tr, tr, tr, noise' sequence indicates a three-time-unit trill call surrounded by noisy silences where a time-unit (50 millisecond) is the window shift of sliding spectral segment inputs for prediction).

To evaluate classification on segment files, containing caller and segmentation information, we discretize the continuous segment intervals of calls and noises into 50-millisecond discrete units for predicted and annotated segment files to get two discrete label sequences of the same size.

We evaluate the hypothesized and reference label sequences (comprising 8 call types and the noise type that indicates silence between calls) by counting correctly and incorrectly classified labels. We calculate the accuracy by
\begin{equation}
  \label{eq:noise_acc}
  noise\_acc = \frac{c_{noise}}{n_{noise}},
\end{equation}

\begin{equation}
  \label{eq:call_acc}
  call\_acc = \frac{c_{call}}{n_{call}},
\end{equation}

\begin{equation}
  \label{eq:total_acc}
  total\_acc = \frac{c_{all}}{n_{all}},
\end{equation}

where $c_{all}$ is the summation of counts of correct noise labels and call labels, $n_{all}$ is the summation number of call and noise labels  (the
sequence length); $c_{noise}$ and  $c_{call}$ the counts of correct noise and call labels; $n_{noise}$ and  $n_{call}$ the total number of noise and call labels.
We calculate the precision by
\begin{equation}
  \label{eq:acc}
  precision = \frac{c_{call}}{c_{call} + e_{noise}},
\end{equation}
where $c_{call}$ is the number of correct call labels (the same call type for the hypothesis and the reference) and  $e_{noise}$ is the number of error noise labels (when predicted as any call but annotated as a noise).
We calculate the recall by
\begin{equation}
  \label{eq:acc}
  recall = \frac{c_{call}}{n_{call}},
\end{equation}
where $n_{call}$ is the number of call labels in annotated reference. And finally, we calculate the F-score by
\begin{equation}
  \label{eq:acc}
  f = \frac{2 * recall * precision}{recall + precision}.
\end{equation}

We also evaluate segmentation with the boundary F-score. The boundary F-score is the harmonic mean of boundary precision and recall, where precision is the number of correct intervals over the number of predicted intervals and recall is the number of correct intervals over the number of annotated intervals. A predicted interval is counted as correct only if there exists one annotated interval whose predicted begin and end times match the annotated begin and end times within a given threshold (which we used a tight threshold of 100 milliseconds compared to our model resolution of 50 milliseconds). Each predicted interval is only allowed to match one reference interval.
\section{Dataset and Experimental setup}
\subsection{Dataset}
\begin{figure}[!ht]
  \centering
  \includegraphics[width=0.9\linewidth]{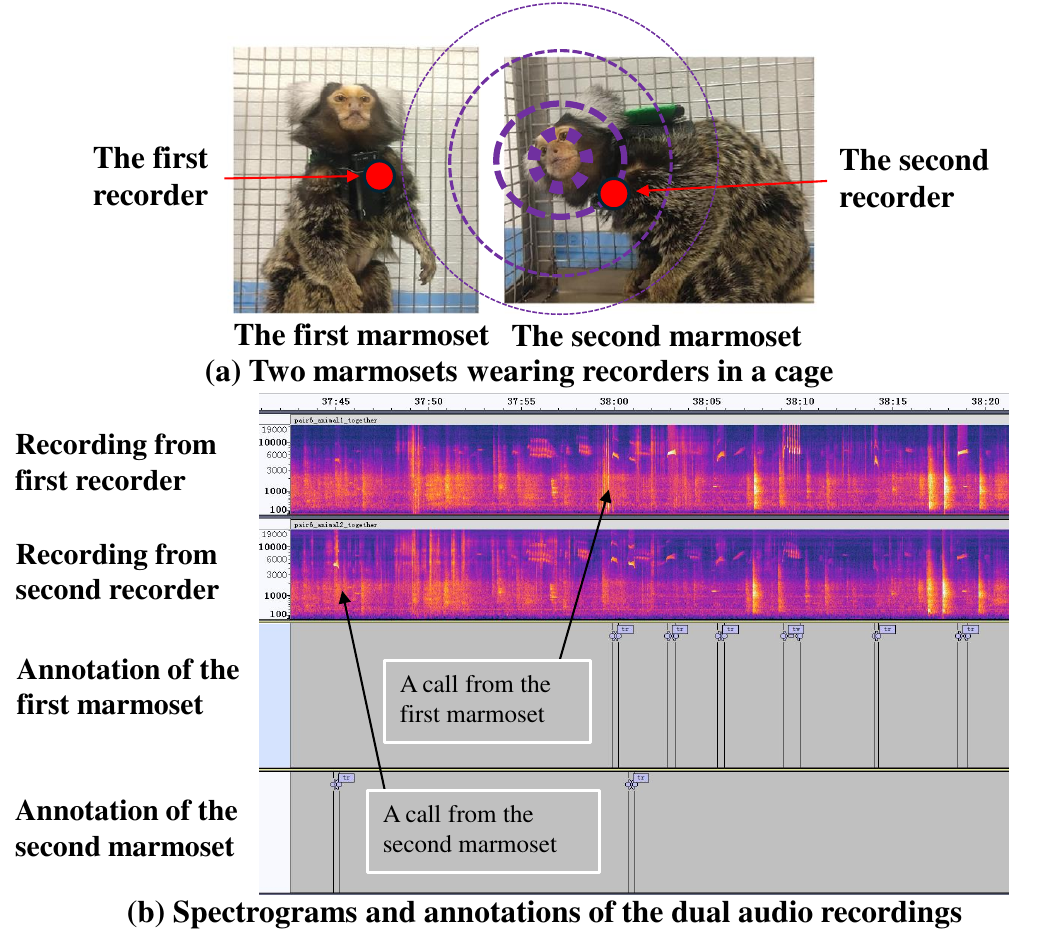}
  \caption{Upper subfigure (a) Two marmosets wearing recorders in a cage. When the second marmoset vocalizes, the first recorder receives a weaker signal than the second one. Marmoset photos from~\cite{landman2020close}.  Lower subfigure (b) Example clip of an annotated dual audio recording.  Calls from other distant cages in the same animal room similarly weak in both audios and are ignored in the annotation~\cite{landman2020close}.}
  \label{fig:dual_audio_takamichi}
\end{figure}

\begin{figure}[t]
  \centering
  \includegraphics[width=0.8\linewidth]{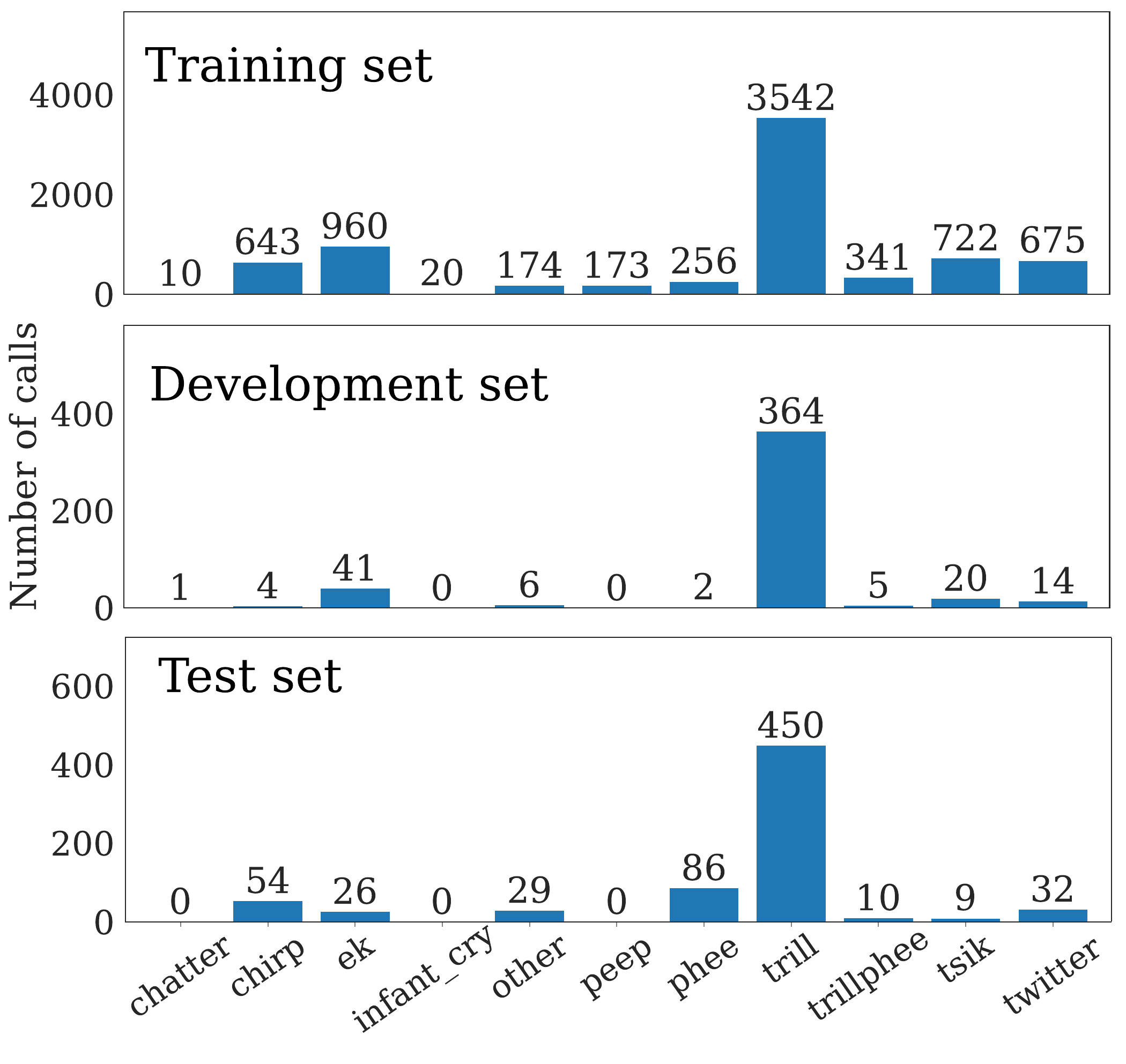}
  \caption{Number of calls in the training, development, and test sets. The dataset~\cite{landman2020close} contains recordings of 10 pairs of marmosets, each pair together in a cage. We use the pair1 as the test set, the pair2 as the development set, and the pair3 to pair10 as the training set.}
  \label{fig:data_division}
\end{figure}

We experimented on the dataset~\cite{landman2020close} designed to study close-range communication between marmosets. We used dual audio recordings from the dataset when two marmosets are together in a cage (excluding recordings when two marmosets are in separate cages). The dataset includes 10 marmoset pairs housed together.


When two marmosets interact each other in a cage while wearing recorders, two simultaneously recorded audios are preserved. The dual-audio recording setup is designed to address the challenging task of call identity annotation. The caller identities were annotated by comparing the spectrograms from the two simultaneous recordings. For example, when the second marmoset makes a call and the first marmoset keeps silent, the recorder worn by the second marmoset should receive a stronger and clearer signal compared to the recorder worn by the first marmoset because the sound wave of the call becomes weaker when it travels further.  The close positioning of recorders to the marmosets' mouths is crucial for this approach. Figure~\ref{fig:dual_audio_takamichi} demonstrates this process and shows  the annotation of call type, call time, and caller identity of a dual audio recording clip.


  \begin{table}[t]
    \centering
    \caption[sys_cmp]{Duration of training, development, and test sets of dual-stream annotated marmoset recordings~\cite{landman2020close} for evaluation of classification, segmentation, and caller identification.}
    \label{tab:data_dur}
    \begin{tabular}{@{}llll@{}}
      \toprule
      \textbf{Data division}  & \textbf{Training set} & \textbf{Development set} & \textbf{Test set} \\ \midrule
      \textbf{Number of pairs} & 8 pairs               & 1 pair                   & 1 pair            \\
      \textbf{Recording time} & 16:07:17              & 1:52:31                  & 2:15:07           \\ \bottomrule
    \end{tabular}
  \end{table}

 \begin{table}[t]
    \centering
    \caption[sys_cmp]{Duration of training, development, and test sets for MAE pretraining on marmoset recordings without annotations.}
    \label{tab:data_dur_mae}
    \begin{tabular}{@{}llll@{}}
      \toprule
      \textbf{Data division}  & \textbf{Training set} & \textbf{Development set} & \textbf{Test set} \\ \midrule
      \textbf{Number of days} & 48 days            & 1 day                 & 1 day          \\
      \textbf{Recording time} & 639:38:19             & 17:45:57                 & 18:38:12          \\ \bottomrule
    \end{tabular}
  \end{table}

\subsection{Data division}
The dataset~\cite{landman2020close} consists of the dual audio recordings of 10 pairs of marmosets, aged from 1.5 to 10 years old, where the 5 female-male pairs are unrelated and the 5 male-male pairs are siblings. The annotations include 11 call types: trill, twitter, chirp, phee, trillphee, other, ek, tsik, chatter, peep, and infant cry (Figure~\ref{fig:data_division}). We divided the dataset with the data of pair1 as the test set, pair2 as the development set, and pair 3 to pair 10 as the training set with duration statistics (Table~\ref{tab:data_dur}) and frequency statistics (Figure~\ref{fig:data_division}). All data come from annotated dual audio recordings of marmoset pairs housed together.

We collected longitudinal recordings for 50 days of a marmoset family with parents and one child in a sound-proof box. The recording is single-stream and we recorded more than 10 hours per day. We pretrained the MAE on the whole dataset (Table~\ref{tab:data_dur_mae}) using 48 days for pretraining after roughly removing some empty audios and long silence periods without vocalization. The audios have no labels and we used MAE self-supervised learning on the marmoset linear spectrogram segments without annotations.

\subsection{Experimental setup for systems}
We built our systems using the identical 8 call types as~\cite{oikarinen2019deep}. We converted the additional rare call types in the dataset~\cite{landman2020close} (other, peep, and infant cry) into the noise type. After adding the caller identity information, our target labels become trill, phee, trillphee, twitter, chirp, tsik, ek, chatter (from the first animal), trill2, phee2, trillphee2, twitter2, chirp2, tsik2, ek2, chatter2 (from the second animal), and noise type (indicating intervals between calls for segmentation).  
  \subsubsection{Our CNN backbone model}
  We implemented our backbone CNN model with the same architecture as~\cite{oikarinen2019deep}. It consists of two CNN streams whose outputs are concatenated and linear-projected for the final prediction. Each CNN stream comprises 4 CNN modules. Each module comprises two identical convolutional layers followed by max-pooling. Whenever max-pooling is applied, number of channels in convolutional layers doubles in subsequent module.
  \subsubsection{Proposed Transformer backbone model}
  We implemented our backbone Transformer model using Vision transformer that divides the original high-resolution $257 \times 256$ linear spectral segments into $16 \times 16$ patches. The spectrogram patches reduce Transformer input lengths compared to pixels and balance the time and frequency resolutions. These patches are linearly embedded and then passed into a Transformer module with a model dimension of 384. The Transformer module consists of 6 blocks, each with 6-head self-attention and linear modules. The linear modules have a hidden dimension of 1536. The outputs of the two Transformer streams are concatenated and passed through a shared linear layer with a dimension of 1024 for the final prediction. We found that larger Transformer models (e.g., model dimension of 768 or deeper layers) would perform poorly on small amounts of available annotated data.

\begin{figure}[t]
  \centering
  \includegraphics[width=\linewidth]{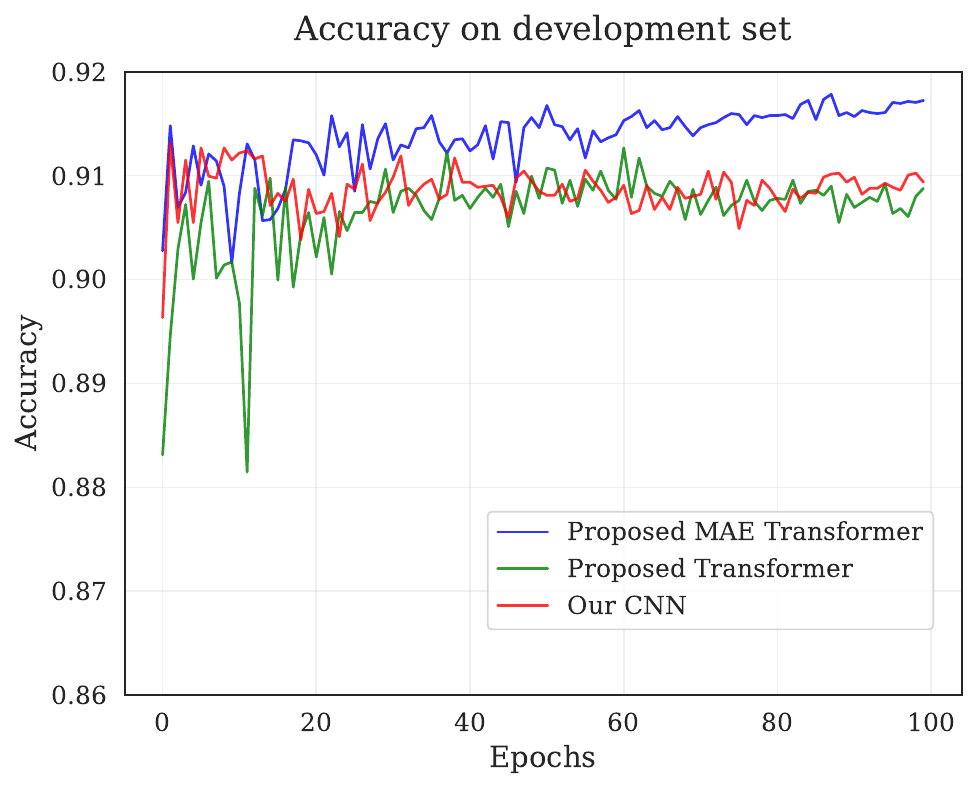}
  \caption{The accuracy of our CNN, proposed Transformer, and MAE Transformer on the development set during training epochs. The proposed MAE Transformer is the Transformer pretrained using a MAE model on marmoset recordings without labels. We observed that the MAE pretrained Transformer showed better training stability and less overfitting.}
  \label{fig:dev_acc}
\end{figure}

  \subsubsection{Proposed Transformer backbone model pretrained by MAE}
  We implemented our backbone Transformer model using an MAE encoder pretrained on the ImageNet training set~\cite{He2022} and 48 days of marmoset spectrogram segments (more than 10 hours per day). We used 12 layers of the pretrained Transformer, with the lower 6 layers frozen used for MAE feature extraction and the upper 6 layers for task-specific fine-tuning (where 6 layers performed the most stably and best among 3, 6, and 12 layers). We initialized identical parameters for the Transformers of both streams. We found that using a slightly larger model with model dimension 768 did not degrade performance on limited annotated data due to good pretrained initialization that stabilized training and MAE features that have good generalization power.

  The pretrained MAE for marmoset recordings uses 12 layers, 12 attention heads, 768 embeddings with a masking ratio of 75\%. The pretraining on the marmoset dataset included 400 training epochs with batch size 256, learning rate 0.00015, 40 warmup epochs, and 0.05 weight decay.

\begin{table}[t]
  \centering
  \setlength{\tabcolsep}{4pt}
  \setlength{\abovecaptionskip}{5pt}
  \setlength{\belowcaptionskip}{0pt}
  \small
  \caption{Separate classification, caller identification, and segmentation evaluations where segmentation evaluation uses the boundary F-score with 100ms tolerance. We used ~\cite{oikarinen2019deep}'s best system as a baseline and evaluated our CNN and proposed Transformer and MAE Transformer systems on the same dataset~\cite{landman2020close}. MAE\_Transformer is the Transformer pretrained using a MAE model on marmoset recordings without labels.}
  \label{tab:result2}
  \vspace{-0mm}
\begin{tabular}{@{}lccc@{}}
\toprule
\textbf{Systems} & \textbf{Call Acc.} & \textbf{Caller Acc.} & \textbf{Boundary F-score} \\
\midrule
\textbf{Baseline} & 0.7212 & 0.7437 & 0.0180 \\
\textbf{CNN} & 0.7575 & 0.7866 & \textbf{0.0182} \\
\textbf{Transformer} & 0.7576 & 0.7867 & 0.0152 \\
\textbf{MAE\_Transformer} & \textbf{0.7811} & \textbf{0.8119} & 0.0166 \\
\bottomrule
\end{tabular}
\vspace{-4mm}
\end{table}

\subsubsection{Proposed CNN/Transformer system with backbone model}
We trained our two-stream system, with CNN or Transformer backbone, to map two $257 \times 256$ 2-dimensional linear spectral segments to the target label. The segments were generated from dual-channel audio using a sliding window of 500ms size with a 150ms shift~\cite{oikarinen2019deep}. The target labeling process was as follows: A call target label was assigned when the middle 150ms part of the segment overlapped with a human-annotated call interval; a noise target label was assigned when the middle 150ms part of the segment did not overlap with any call annotations. These call and noise label targets enable the model to classify and segment long-hour recordings.

  To handle long inactive silent intervals in long-hour recordings when marmosets do not vocalize, we implemented two strategies. First, we randomly discarded 4/5 of noise segments to improve training efficiency and balance the dataset. Second, we applied data augmentation by randomly roll-shifting each spectral segment 1-5 pixels vertically and horizontally during model training. These approaches addressed the uneven distribution of vocalizations while enhancing the system's ability to accurately classify marmoset calls.
  
  To use the system to monitor marmoset behavior for long-hour recordings, we applied a streaming technique to enhance model prediction speed. This process involves two main steps: First, we split the test audios into non-overlapping 2500ms segments. Second, we created 50 sub-segments from each 2500ms segment, using a window size of 500ms and a window shift of 50ms: the first 41 sub-segments are complete within the current segment, while the last 9 concatenate parts from the next 2500ms segment. During prediction, after feeding a batch into the model, we obtain predicted 50ms intervals corresponding to 50 subsegments in the batch. This approach, inspired by~\cite{oikarinen2019deep}, allows for efficient processing of long-duration audio recordings. We also implemented spectral feature extraction with Pytorch that uses GPU for better efficiency.

  We implemented our Transformer system using a Vision Transformer model. Our two-stream Transformer system maintains the same overall architecture as our CNN system but replaces the CNN backbone with a Vision Transformer model. Following~\cite{oikarinen2019deep}, our system uses Adam optimizer with a learning rate of 0.0003, decaying by a factor of 0.97 each epoch.

\begin{table}[t]
\centering
\caption{Results of our CNN and proposed Transformer and MAE Transformer systems and~\cite{oikarinen2019deep}'s best system as a baseline on the same dataset~\cite{landman2020close}.  The proposed MAE Transformer is the Transformer pretrained using a MAE model on marmoset recordings without labels.}
\label{tab:result}
\setlength{\tabcolsep}{4pt}
\small
\begin{tabular}{@{}lccc@{}}
\toprule
\textbf{Systems} & \textbf{F-score} & \textbf{Recall} & \textbf{Prec.} \\
\midrule
\textbf{Baseline}  & 0.7686 & 0.7212 & 0.8227 \\
\textbf{Our CNN system} & 0.7901 & 0.7575 & 0.8257 \\
\textbf{Proposed Transformer} & 0.7940 & 0.7576 & \textbf{0.8341} \\
\textbf{Proposed MAE Transformer} & \textbf{0.7998} & \textbf{0.7811} & 0.8195 \\
\bottomrule
\end{tabular}
\vspace{2mm}
\vspace{2mm}
\begin{tabular}{@{}lccc@{}}
\toprule
\textbf{Systems} & \textbf{Total Acc.} & \textbf{Noise Acc.} & \textbf{Call Acc.} \\
\midrule
\textbf{Baseline} & 0.9899 & 0.9963 & 0.7212 \\
\textbf{Our CNN system} & 0.9907 & 0.9962 & 0.7575 \\
\textbf{Proposed Transformer} & 0.9909 & \textbf{0.9964} & 0.7576 \\
\textbf{Proposed MAE Transformer} & \textbf{0.9909} & 0.9959 & \textbf{0.7811} \\
\bottomrule
\end{tabular}
\end{table}



\section{Result and discussion}
We applied our system directly to each raw long-hour audio recording (approximately 2 to 3 hours) without extra processing. We compare our PyTorch-implemented two-stream CNN and transformer system with the best sytem of~\cite{oikarinen2019deep} (open-sourced) on the same dataset~\cite{landman2020close} of annotated dual audio recordings with the data division shown in the Figure~\ref{fig:data_division}.

The Figure~\ref{fig:dev_acc} shows real-time accuracy over 100 epochs on the development set during training of the CNN, Transformer, and MAE pretrained Transformer where MAE is pretrained with marmoset recordings without annotation. We observed that: 1) For CNN, the small model quickly converges on the limited training data within 25 epochs and then overfits without recovery. 2) For the Transformer model, overfitting occurs much later around 60 epochs, but its starting accuracy is low and shows a steep drop in accuracy due to unstable training. 3) For the MAE pretrained Transformer (6 frozen layers for MAE features and 6 layers for fine-tuning for MAE Transformer vs. 6-layer Transformer), the starting accuracy after the first epoch is even better than CNN. We did not observe any overfitting during fine-tuning, with accuracy continuing to increase until the end of 100 epochs. The MAE pretrained Transformer outperformed both the Transformer and CNN by a large margin.

Our proposed Transformer systems surpass the CNN system and baseline system~\cite{oikarinen2019deep} across all F-score and total accuracy (Table~\ref{tab:result}); the bootstrap one-tailed unpaired t-test on F-score comparing our CNN and MAE ViT is significant at the 0.05 level. Our proposed Transformer systems also achieve the best performance in separate evaluations of classification, and caller identification (Table~\ref{tab:result2}).

      The Transformer performs best using global contextual modeling to better capture overall call patterns. However, the Vision Transformer patchizes input spectrograms, resulting in lower model resolution than the CNN model. The transformer prediction intervals are usually wider than annotated truth, penalized by strict segmentation evaluation such as 100ms tolerance boundary F-score (Table~\ref{tab:result2}); we also observed penalty at 50ms and 200ms tolerance boundary F-scores. In future work, we will improve Transformer system for better resolution modeling.

      Our systems can be applied to raw, long-hour audio recordings to segment and classify calls, as well as identify callers. This capability provides a valuable tool for recording marmoset interactions, which can support future studies on social behavior, development, and abnormalities in marmoset vocalizations. Such research could offer insights into communication, evolution, and dysfunction of the vocal language of marmoset. The research also helps facilitate comparisons between marmoset and human infant vocal development in family environments. Additionally, the Transformer update of our system offers potential for developing a unified multimodal model that integrates both video and audio inputs.

\section{Conclusion}
We proposed self-supervised pretraining that enables robust marmoset call segmentation, classification, and caller identification with two-stream Transformer systems. The Transformer systems outperform previous CNN approaches. Our Transformer pretrained by MAE on hundreds of hours of marmoset recordings without annotations shows the best performance with more stable training and less overfitting compared to CNN and Transformer without pretraining. Our system efficiently processes long-hour or full-day marmoset recordings for vocal communication between marmosets in spontaneous interactions, advancing research on language evolution, development, and dysfunction.


\section{Acknowledgements}
We would like to thank Yingjie WANG for early exploration in models. Part of this work was supported by RIKEN Pioneer Project 22.

\bibliographystyle{IEEEtran}
\bibliography{mybib}

\end{document}